\documentclass[aps,prl,preprint,showpacs]{revtex4}
\usepackage{xcolor}
\usepackage[hypertex,backref]{hyperref}
\usepackage{graphicx,subfigure}
\usepackage{amsmath,amsfonts,amssymb}

\newcommand{\f}{\frac}
\newcommand{\suml}{\sum\limits}
\newcommand{\prodl}{\prod\limits}
\newcommand{\m}{\mathbf}

\newcommand{\oo}{\infty}
\graphicspath{{D:/localtexmf/Mytex/LGH/eqstcan/mathplots/}}
\begin{document}
\title[Global isomorphism]{Global isomorphism between the
Lenard-Jones fluids and the Ising model}
\author{V.L. Kulinskii}
\email{kulinskij@onu.edu.ua}
\affiliation{Department for Theoretical
Physics, Odessa National University, Dvoryanskaya 2, 65026 Odessa, Ukraine}
\begin{abstract}
The interpretation of the linear character of the observable
classic rectilinear diameter law and the linear character of
the Zeno-line (unit compressibility line $Z=1$) on the basis of global isomorphism between Ising model (Lattice Gas) and simple fluid is proposed. The correct definition of the limiting nontrivial Zeno state is given and its relation with the locus of the critical point is derived within this approach. We show that the liquid-vapor part of the phase diagram of the
molecular fluids can be described as the isomorphic image of
the phase diagram of the Lattice Gas.
It is shown how the the position of the
critical points of the fluids of the Lenard-Jones type can be
determined basing on the scaling symmetry. As a sequence the
explanation of the well known fact about ``global`` cubic
character of the coexistence curve of the molecular fluids is
proposed.
\end{abstract}
 \maketitle
\section{Introduction}\label{sec_intro}
Since the seminal doctoral thesis of van der Waals \cite{vdW_thesis} where the Principle of the Corresponding States was formulated the search for the unifying principles for the description of the variety of thermodynamical properties of complex matter is the key point of statistical physics. By the rigorous methods of the Statistical Mechanics the universal character of the van der Waals (vdW) equation was demonstrated and the numerous extension were proposed \cite{book_hansenmcdonald}.
The consequences derived from the vdW equation and the observed deviations became the driving force for the further theoretical developments. One of the main achievement of van der Waals was the description of the phase liquid-gas equilibrium which terminated at the corresponding critical point (CP). It opened the possibility to connect the observable thermodynamic quantities with the characteristics of intermolecular interactions.
In the modern theory of the Critical Phenomena the ideology of the isomorphism classes of the critical behavior provides the description of the real systems using the results obtained for the model systems like lattice models \cite{book_baxterexact}. In particular the molecular liquids with short range interactions of the Lenard-Jones (LJ) type belong to the isomorphism class of the Ising model. The latter can be easily mapped to the Lattice Gas (LG) model,
which is determined by the Hamiltonian:
\begin{equation}\label{ham_latticegas}
  H = -J\suml_{
\left\langle\, ij \,\right\rangle
  } \, n_{i}\,n_{j} - \mu \,\suml_{i}\,n_{i}\,,
\end{equation}
where $n_{i} = 0,1$ whether the site is empty or occupied correspondingly. The quantity $J$ is the energy of the site-site interaction of the nearest sites $i$ and $j$, $\mu$ is the chemical potential. The temperature variable corresponding to the Hamiltonian \eqref{ham_latticegas} is denoted as $t$. The order parameter is the probability of occupation of the lattice site $x = \left\langle\, n_i
\,\right\rangle$ and serves as the analog of the density. The
phase diagram of the LG is symmetrical with respect to the line $x_{0} = 1/2$ and formally extends up to the low temperature region $t\to 0$, where the limiting states $x=0$ and $x = 1$ exist only.

Two of the consequences of the vdW equation are of primary
importance for our consideration. The first one is the
well-known rectilinear law for the diameter
of the coexistence curve in terms of the particle density $n$ and the thermodynamic temperature $T$:
\begin{equation}\label{densdiam}
  n_{d} = \f{n_{liq}+n_{gas}}{2\,n_c} = 1+ A\,\left|\tau \right|+\ldots\,,\quad \tau  = \f{T-T_c}{T_c}
\end{equation}
where $n_{i}, \, i = liq,\,gas$ are the values in the corresponding phases, $n_c,T_c$ are the critical density and the temperature. The second fact is another linear law which was derived by Batchinsky \cite{eos_zenobatschinski_annphys1906} long ago from the classical vdW equation:
\begin{equation}\label{p_vdw}
  P = \f{n\,T}{1-n\,b}-a\,n^2\,.
\end{equation}
The curve defined by the equation $Z = P/(n\,T) =1$ for the compressibility factor $Z$ as it directly follows from Eq.~\eqref{p_vdw} is a straight line:
\begin{equation}\label{vdw_z1}
  \f{n}{1/b}+\f{T}{T^{\text{(vdW)}}_B} = 1\,.
\end{equation}
where $T^{\text{(vdW)}}_B = a/b$ is the Boyle temperature determined by the vdW constants in accordance with the virial expansion for Eq.~\eqref{p_vdw}.
In work of Ben-Amotz and Herschbach \cite{eos_zenobenamotz_isrchemphysj1990} the line $Z = 1$ is called by the Zeno-line.

Both relations \eqref{densdiam} and \eqref{z1} are approximate and have phenomenological character. But surprisingly they are
observed for a wide variety of fluids. As was noted in  \cite{pcs_guggenheim_jcp1945} the law \eqref{densdiam} is fulfilled  in rather broad temperature interval of the vapor-liquid phase coexistence region $0.65\lesssim T/T_c\lesssim 1$. In the vicinity of the CP the deviations from the linearity become noticeable \cite{crit_diamermin1_prl1971,crit_rehrmermin_pra1973}.

In series of works \cite{eos_zenoapfelbaum_jchempa2008,eos_zenoapfelbaum_jcp2004,
eos_zenoapfelbaum_jchemp2006,zenoline_potentials_jcp2009,
eos_zenoapfelbaum1_jpcb2009} the performed analysis of the data for both the real fluids and the systems with model potentials discovered a number of interesting relations between coordinates of the CP, the law of the rectilinear diameter \eqref{densdiam} and the Zeno-line:
\begin{equation}\label{z1}
  \f{n}{n_b}+\f{T}{T_B} = 1\,.
\end{equation}
Here $T_B$ is the Boyle temperature and for $n_b$ the following equation followed from the virial expansion was proposed in \cite{eos_zenoapfelbaum_jchempb2008}:
\begin{equation}\label{nb_apfelbaum}
T_B = n_b\,\f{B_3(T_B)}{B'_2(T_B)}\,.
\end{equation}
The phenomenological concept of the ``Triangle of Liquid-Gas States`` has been formulated in
\cite{eos_zenoapfelbaum_jchemp2006}. It is heavily based
on the cute observation that the Zeno-line is the tangent
to the binodal extrapolated to low temperature region.
Then the liquid-vapor binodal is inscribed into triangle
formed by the straight lines: the coordinate axes $n$ and
$T$ and the Zeno-line.

Obviously, the thermodynamic states which satisfy the equation $Z=1$ include the trivial states with negligible density $n\to 0$. This is nothing but the coordinate axis on the thermodynamic $n-T$ plane. Thus the set of the Zeno states is formed by the union of two lines $n=0$ and \eqref{z1}. The statement \eqref{z1} about the linear nature of Zeno-states is far from trivial. Indeed the expression for $Z$, which for the system with binary potential interaction $\Phi(r_{12})$ reads as (see e.g. \cite{book_hansenmcdonald}):
\begin{equation}\label{compressfactor}
Z =\f{P}{n\,T} = 1-\f{2\pi \,n}{3\,T}\int r^3
\frac{\partial\, \Phi(r)}{\partial\, r}\,g_2(r;n,T)\, d\,r\,,
\end{equation}
where $g_2$ is the pair correlation function. Therefore the linearity of the Zeno states is due to quite specific structure of the dependence of correlation function $g_2(r;T,n)$ on the thermodynamic state.
%

Below, following the arguments of previous work
\cite{eos_zenome_jphyschem2010}, we show that these facts can
be casted into elegant geometrical formulation and expand them into general case of the short range power-like attractive
potentials in $d$ dimensions. We show that the locus of the CP can be estimated using the correspondence between the scaling properties of the Hamiltonians for the lattice gas and the fluids with the power-like interactive potentials.

\section{The Zeno-line and the global isomorphism}\label{sec_isomorphzeno}
As was shown in \cite{eos_zenome_jphyschem2010} the linearities of \eqref{densdiam} and \eqref{z1} can be derived on the basis of the assumption of the existence of the
global isomorphism of the real liquid-gas part of the phase
diagram of the lattice gas model. Note that the line $x=1$ of
the LG can be thought of as the analog of the Zeno-line. Indeed the pair correlation function $g_2(r)$ of the LG vanishes
identically for such ``holeless`` states according to the
definition. The same is true for the state with empty sites
$x=0$. Thus the line $x= 0$ can be identified with the zeroth
density axes $n=0$ of the real fluid, while the line $x= 1$,
obviously, can be identified with the Zeno-line $Z=1$. To conserve the linearity and the adjacency
properties of the characteristic elements the isomorphism between LG and the fluid
should be chosen in the class of the projective mappings
\cite{hartshorn_projgeom} and has the form:
\begin{equation}\label{projtransfr_nx}
  n =\, n_b\,\f{x}{1+z \,t}\,,\quad
  T =\, T_Z\,\f{z\, t}{1+z \,t}\,,
\end{equation}
where
\begin{equation}\label{alpha}
z  = \f{T_c}{T_Z-T_c}\,.
\end{equation}
The coordinates of the CP for the liquid are:
\begin{equation}\label{cp_fluid}
  n_{c} = \f{n_b}{2\left(\,1+z\,\right)}\,,\quad   T_{c} = T_Z\, \f{z}{1+z}\,,
\end{equation}
provided that the LG temperature variable $t$ is scaled so that $t_c=1$. In view of Eq.~\eqref{alpha} it should be noted that if $T_Z$ and $n_b$ are fixed then the parameter $z$ parameterizes the locus of the CP:
\begin{equation}\label{cpline}
  \f{n_c}{n_b} +   \f{T_c}{T_Z} = \f{z +1/2}{1 +z}
\end{equation}
This means that for the substances belonging to the same class of the corresponding states the loci of the critical points  scaled to the $T_Z$ and $n_b$ lie along the straight line. This correlates with the empirical analysis in \cite{eos_zenoapfelbaum_jchempb2008}. Because both $T_Z$ and $T_c$ are determined by the interparticle potential the parameter $z$ can be connected with its symmetry properties. Note that all results obtained rely solely on the geometrical facts about phase diagrams of the Lattice Gas and the fluid and do not depend on the specific details of the interactions.

Two limiting Zeno states $T\to 0,\, n\to n_b$ and $T\to T_Z,\,n\to 0$ form the triangle of the Zeno states \cite{eos_zenoapfelbaum_jchemp2006}. It should be noted that the temperature $T_Z$ of the limiting Zeno state with $n_Z\to 0$ does not necessary coincide with the Boyle temperature $T_B$ as was originally proposed in \cite{eos_zenoapfelbaum_jchempa2008}. In \cite{eos_zenoapfelbaum_jchempa2008} the Boyle state $n\to 0\,, T=T_B$ as an \textit{obvious} limiting Zeno state is used and then the parameter $n_b$ is defined by Eq.~\eqref{nb_apfelbaum}. But any state on $n$-axes can be used in such way. Two limiting Zeno states $n\to 0\,,T=T_Z$ and $n=n_b\,,T\to 0$ belong to two physically different thermodynamic regions. Therefore there is no \textit{apriori} physical reasons to define the value $n_b$ basing on $T_B$. From the point of view of the virial expansion:
\begin{equation}\label{virial}
  Z = \f{P}{n\,T} = 1+\suml_{k=1}^{\oo} \,B_{k+1}(T)\,n^{k}
\end{equation}
where $B_{k}(T)$ is the virial coefficients \cite{book_balesku_statmech}, the limiting state $T_Z$ with $n_Z\to 0$ is the point where the nontrivial branch of the solution of $Z=1$ emerges from the axis $n=0$ with the specific tangent determined by $T_Z$ itself. Clearly the Boyle temperature $T_B$ determined by the condition:
\begin{equation}\label{tboyle}
  B_2(T_B)=0\,,
\end{equation}
is the simplest choice but in general $T_Z \ne T_B$ because \eqref{tboyle} does not guarantee the correct slope of the Zeno line. Indeed if one search the solution of the equation:
\begin{equation}\label{virialexp}
Z-1 = n\,B_2(T)+B_{3}(T)\,n^2+\ldots = 0
\end{equation}
in a form of \begin{equation}\label{zline}
  \f{n_Z}{n_b} = 1- \f{T}{T_Z}
\end{equation}
in the vicinity of $T=T_Z$ then obviously the infinitesimal condition:
\begin{equation}\label{dz}
\f{dZ(n_Z(T),T)}{dT} = \left.  \frac{\partial\,Z}{\partial\, n}\right|_{T\to T_Z,\,n\to 0}\,\f{dn_Z}{dT} +   \left.\frac{\partial\, Z}{\partial\, T}\right|_{T\to T_Z,\,n\to 0} = B'_2(T) + B_3\,\f{dn}{dT} = 0
\end{equation}
should be fulfilled. Using Eq.~\eqref{virial} we get that $T_Z$ is the root of the equation:
\begin{equation}\label{TZ}
T_Z = n_b\,\f{B_3(T_Z)}{B'_2(T_Z)}\,.
\end{equation}
This is the rigorous definition of the limiting Zeno state at
$n\to 0$ from which the Zeno line emerges. Note that for vdW equation with $B_2 = b-\f{a}{T}$, $B_3 = b^2$ and $n_b = 1/b$, Eq.~\eqref{TZ} gives the standard result $T_Z = a/b$. Therefore this is the specific property of the vdW equation that the temperatures $T_Z$ and $T_B$ coincide. Actually the value
$T_Z$ depends on the dimensionless packing parameter $\eta_b = n_b\,\sigma^d$ in $d$ dimensions, where $\sigma$ is the characteristic scale of the potential, e.g. the hard core diameter. This parameter is determined for the opposite limiting Zeno state with $T\to 0$.
Though Eq.~\eqref{TZ} formally coincide
with Eq.~\eqref{nb_apfelbaum}, the approach proposed for the determination of the limiting Zeno-state at $n\to 0$ is  different from that used in \cite{eos_zenoapfelbaum_jchempa2008}. Thus the situation is inverse: rather the value $n_b$ determines the value $T_Z$ and not vice versa. In order to make this consistently with the definition of the Zeno-line we choose $n_b$ so that $T_Z = T^{\text{(vdW)}}_B$, where $T^{(vdW)}_B = a/b$ is the Boyle temperature in the van der Waals approximation. We use this to calculate the locus of the critical point in Section~\ref{sec_cplocus}. Using the standard expressions for the virial coefficients $B_{k}$ for Lenard-Jones potential
\begin{equation}\label{lj612}
  \Phi(r) = -4U_0\left(\,\left(\,\f{\sigma}{r}\,\right)^6 - \left(\,\f{\sigma}{r}\,\right)^{12}\,\right)\,.
\end{equation}
the solution of \eqref{TZ} for 3-dimensional system gives $n_b\,\sigma^3  \approx 0.965$ and
$T_Z = 4$ while $T_B\approx 3.42$.
\section{Scale invariant mean-field theory}\label{sec_scaleinvmf}
In this section using the proposed isomorphism we propose the variant of the mean-field approach to calculate the locus of the critical point. It exploits explicitly the scale symmetry which  inherent to the power-like interaction potentials. We call it scale-invariant mean-field theory.

Let us show that the parameter $z$ can be determined if the attractive part of the interaction potential possesses the scaling symmetry. The starting point is that using \eqref{cp_fluid} it is easy to derive the following relation:
\begin{equation}\label{cp_scalingalpha}
- \f{d\,\ln{\left(\,T_c/T_Z\,\right)}}{d\,\ln{\left(\,n_c/n_b\,\right)}} = \f{1}{z}
\end{equation}
The locus of the CP is determined mostly by the
attractive part of the interaction potential and the size of
the particles (see e.g. \cite{book_balesku_statmech}). These parameters can be connected with the
corresponding ones for the LG model. Suppose that  the potential of the $d$-dimensional system is $\Phi(r) = \Phi_{\text{rep}}(r)+\Phi_{\text{attr}}(r)$, where $\Phi_{\text{rep}}, \Phi_{\text{attr}}$ are the repulsive and the attractive part correspondingly. Additionally we assume that the attractive part has the power-like behavior $\Phi_{\text{rep}}(r)\sim r^{-(d+\varepsilon)}\,,\varepsilon >
0$. Then the corresponding energy of the interaction for the configuration of the number density $n(\m{r})$:
\begin{equation}\label{ham_int}
  E_{\text{int}} = \f{V}{2}\,\int \Phi(r_{12})\,n(\m{r}_{1})\,n(\m{r}_{2})\,d\m{r}_{12}
\end{equation}
can be compared with the LG Hamiltonian \eqref{ham_latticegas}. Now we put the constraint of the scale invariance of the partition function:
\begin{equation}\label{partfunc}
  \mathcal{Z} = \int \,e^{-E_{\text{int}}/T}\,d\Gamma \,,\quad d\Gamma = \prodl_{i=1}^{N}\,d\m{r}_{i}\,,
\end{equation}
corresponding to \eqref{ham_int} with respect to scaling transformations of $n\to \lambda^{\Delta_{n}}\,n$ and $T\to \lambda^{\Delta_{T}}\,T$, where $\lambda$ is the scale parameter. The exponents $\Delta_{n}$ and $\Delta_{T}$ can be found using standard similarity considerations widely used in condensed matter theory (see e.g. \cite{crit_scalingsmirnov_ufn2001en}). E.g. for the coordinates of the CP we have:
\begin{equation}\label{cp_simple}
n_c\sim \f{1}{r^d_c}\,,\quad T_c \sim \Phi_{\text{attr}}(r_c) = \Phi_{\text{attr}}\left(n^{-1/d}_c\right)\,,
\end{equation}
where $r_c$ is the mean interparticle distance in the critical state. Therefore from \eqref{cp_scalingalpha} and \eqref{cp_simple} we can write the relation:
\begin{equation}\label{tcphi}
  \f{d\,\ln \left(\,T_c/T_Z \,\right) }{d\,\ln n_c/n_b} =   \f{d\,\ln \Phi_{\text{attr}}(n^{-1/d}_c) }{d\,\ln n_c/n_b}
\end{equation}
which gives the following consistency relation:
\begin{equation}\label{alpha_d}
 1/z  =1+\varepsilon/d\,.
\end{equation}

Thus we obtain the one-parameter group of scaling transformations for the locus of the CP connected with the change of the characteristic scale of the interaction:
\begin{equation}\label{lgcp_scaling}
n_c(z)\to \,n_c(0)\,\lambda^{-1}\,,\quad   T_c(z) \to T_c(0)\,\lambda^{\left(\, 1+\varepsilon/d\,\right)}\,.
\end{equation}
Eq.~\eqref{lgcp_scaling} is equivalent to the constraint:
\begin{equation}\label{cp_inv}
 n_c(\lambda)^{1+\varepsilon/d}\,T_c(\lambda) = const\,.
\end{equation}
%

Let us show that Eq.~\eqref{alpha_d} can be considered as some
consistency condition between the scaling of for the LG Hamiltonian
\eqref{ham_latticegas} and that for \eqref{ham_int}. Note that
Eq.~\eqref{lgcp_scaling} represents in particular the scaling of the
molecular size $n_b$ and corresponding scaling of energy scale given
by $T_B$ or equivalently $T_Z$. To connect the LG model with the LJ
fluid where the interaction has the form \eqref{ham_int} we take into account that under the scale transformation the loci of the
corresponding critical points of these systems should be scaled
consistently. Indeed, according to the structure of the LG Hamiltonian \eqref{ham_latticegas} the analog of Eq.~\eqref{cp_inv} for the critical
point of the LG is as following:
\begin{equation}\label{scaleinv_lg}
x^2_c\,t_c = const \Leftrightarrow  \f{d\,t_c}{d\,x_c} = - 2\,\f{t_c}{x_c}\,.
\end{equation}
Then the consistency condition between the scaling of the critical point coordinates of the LJ fluid \eqref{cp_scalingalpha} and the Lattice Gas \eqref{scaleinv_lg} is:
\begin{equation}\label{consistency}
\f{T_c/T_Z}{n_c/n_b}\,\left(\,1+\f{\varepsilon}{d}\,\right)  = 2 \,\quad \Rightarrow \quad \f{1}{z} = 1+\f{\varepsilon}{d}
\end{equation}
%

In accordance with \eqref{projtransfr_nx} we get:
\begin{equation}\label{tcnc}
  \f{T_c}{T_Z} = \,\f{1}{2+\f{\varepsilon}{d}}\,,\quad \f{n_c}{n_b} = \,\f{1+ \f{\varepsilon}{d}}{2\left(\,2+ \f{\varepsilon}{d}\,\right)}\,.
\end{equation}
The case of the attractive potential of the van der Waals  forces in $d$-dimensions corresponds to $\epsilon =d\,,\,z = 1/2$ and we get the estimates:
\begin{equation}\label{tcnc_potential}
  T_c = T_Z/3\,,\quad n_c = n_b/3\,.
\end{equation}
In the following Section we use \eqref{tcnc}
to calculate the locus of the critical point
of the fluid where the interaction potential
is of the LJ type \eqref{lj612}.
\section{The locus of the critical point for the Lenard-Jones fluid}\label{sec_cplocus}
To test the predictions \eqref{tcnc} we compare them with the
results of the computer simulations for the potential
\eqref{lj612} for $2D$, $4D$ and $5D$ dimension geometries
available in \cite{crit_lj2dim_jcp1990,crit_lj4dim_jcp1999}.

In accordance with the results of Section~\ref{sec_isomorphzeno} we determine the packing parameter $\eta_b = n_b\,\sigma^d$ so that:
\begin{equation}\label{tb}
T_Z = T^{(vdW)}_B = a/b = \f{4d}{6 - d}\,,\quad a = 2^{d-1}\,\f{4d}{6 - d}\,,\quad b = 2^{d-1}\,,
\end{equation}
where the particle volume $v_{d}(\sigma)$ and $U_0$ are put to units.
The result are placed in Table~\ref{tab_tztb}.

\begin{figure}
{\includegraphics[scale=0.75]{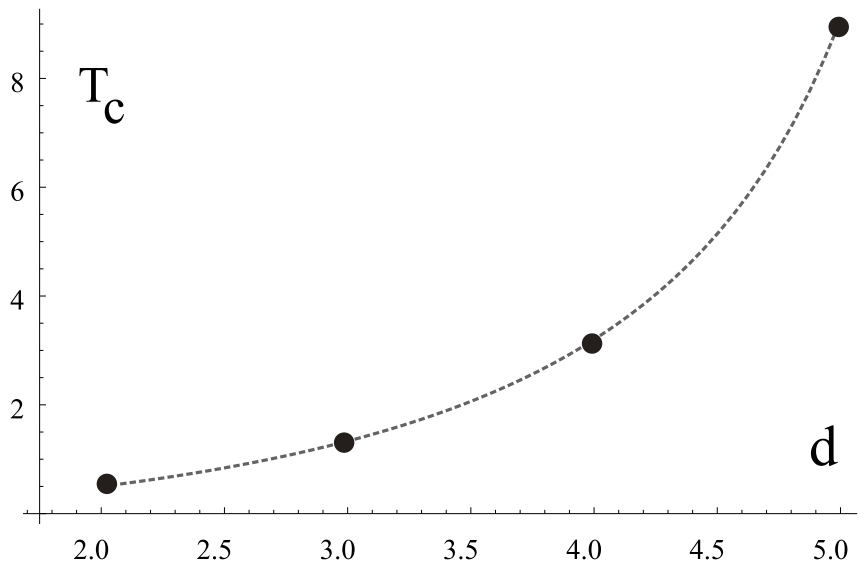}}
\caption{Critical temperature $T_c$ for the potential \eqref{lj612} of the $d$-dimensional LJ fluid according to Eq.~\eqref{tcnc}.}\label{fig_tclj}
\end{figure}

\begin{figure}
\includegraphics[scale=0.75]{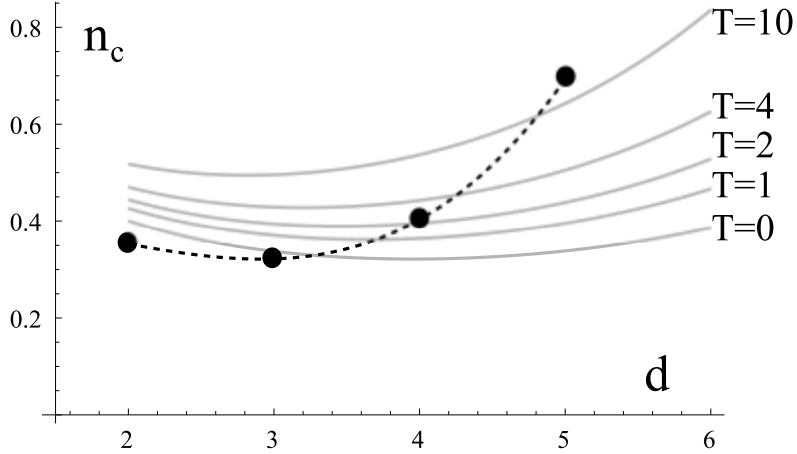}
\caption{Critical density $n_c$ for the potential \eqref{lj612} of the $d$-dimensional according to Eq.~\eqref{tcnc} (dashed line is the guide for an eye) and the packing fraction $\eta(T,d)$ given by Eq.~\eqref{packing} (grey lines) .}\label{fig_nclj}
\end{figure}

The locus of the CP is obtained easily according to
\eqref{tcnc} and \eqref{tb} (see Fig.~\ref{fig_tclj},\ref{fig_nclj}). The comparison with the results of the simulations is in Table~\ref{tab_tcnc}. Note that expressions \eqref{tcnc} and \eqref{tb} allow to clarify the
fact noted in \cite{crit_lj4dim_jcp1999} about different
characters of the dependencies of $T_c$ and $n_c$ on the
dimensionality $d$. The value $T_c$ strongly depends on the
dimension because of the dependence of $T_B$ while the
$d$-dependence of $n_c$ is rather weak.

Note that nonmonotonic dependence of the critical density on the number of dimensions $d$ can be interpreted in terms of the proposed isomorphism as follows. The equilibrium interparticle distance for the LJ potential \eqref{lj612} is $r_0 = \sigma\,2^{1/6}$ and can be considered as the spacing of the cubic lattice for the LG. The effective radius $r_{\text{eff}}$  of excluded volume occupied by the particle from the physical point of view can be defined by the obvious energetic condition:
\begin{equation}\label{reff}
  \Phi(2r_{\text{eff}}) = T\quad  \Rightarrow \quad r_{\text{eff}}(T)\,.
\end{equation}
So the packing of spheres of radii $r_{\text{eff}}$ on the corresponding cubic lattice can be found as
\begin{equation}\label{packing}
  \eta(T,d) = \f{2^{d/6}\,\sigma^d}{V(d)\,r_{\text{eff}}(T)^d}\,,
\end{equation}
where $V(d) = 2\f{\pi^{d/2}}{\Gamma(\f{d}{2})}$\,\, is the volume of the unit ball in $d$ dimensions. Obviously $n_c$
depends on $\eta(T,d)$ monotonically. The dependence of $\eta(T,d)$
on the dimension $d$  is shown on Fig.~\ref{fig_nclj} by grey lines and
shows the minimum in the interval $d\le 4$ in dependence on the temperature.
Note that at $T=0$, when $r_{\text{eff}} = \sigma $
this minimum reaches exactly at $d=4$.
\begin{table}
  \centering
  \begin{tabular}{|c|c|c|c|c|}
    \hline
LJ ``6-12`` fluid& \hspace{0.2cm} 2D \hspace{0.2cm} &\hspace{0.2cm} 3D \hspace{0.2cm} &\hspace{0.2cm} 4D \hspace{0.2cm}&\hspace{0.2cm} 5D \hspace{0.2cm}\\
        \hline
    $T_Z$ &2 & 4& 8& 20\\
    \hline
    $n_b\,\sigma^d$ &0.91 & 0.965& 1.342& 2.54\\
    \hline
    $T_B$ &1.56& 3.418& 9.01 & 40.4 \\
    \hline
  \end{tabular}
  \caption{The results of computation of $T_Z$ and $T_B$  for LJ potential \eqref{lj612} for different dimensions and the packing factor $n_b\,\sigma^d$.}\label{tab_tztb}
\end{table}

\begin{table}
  \centering
  \begin{tabular}{|c|c|c|c|c|}
    \hline
LJ ``6-12`` fluid & \hspace{0.2cm} 2D \hspace{0.2cm} &\hspace{0.2cm} 3D \hspace{0.2cm} &\hspace{0.2cm} 4D \hspace{0.2cm}&\hspace{0.2cm} 5D \hspace{0.2cm}\\
        \hline
    $T_c$ &0.5 & 1.333& 3.2& 9.08\\
    \hline
    $T^{(num)}_c$, \cite{crit_lj4dim_jcp1999} &0.515& 1.312& 3.404 & 8.8 (?) \\
    \hline
    $n_c$ &0.353& 0.322& 0.404& 0.693\\
    \hline
    $n^{(num)}_c$, \cite{crit_lj4dim_jcp1999} &0.355 & 0.316 & 0.34 & - \\
    \hline
  \end{tabular}
  \caption{The comparison of the estimates \eqref{tcnc} with the results of numerical simulations \cite{crit_lj4dim_jcp1999}. The value $T^{5D}_c\approx 8.8$ of the critical temperature for $5D$ case was obtained in \cite{crit_lj4dim_jcp1999} by simple extrapolation of the results for lower dimensions.}\label{tab_tcnc}
\end{table}
\begin{figure}
\includegraphics[scale=0.4]{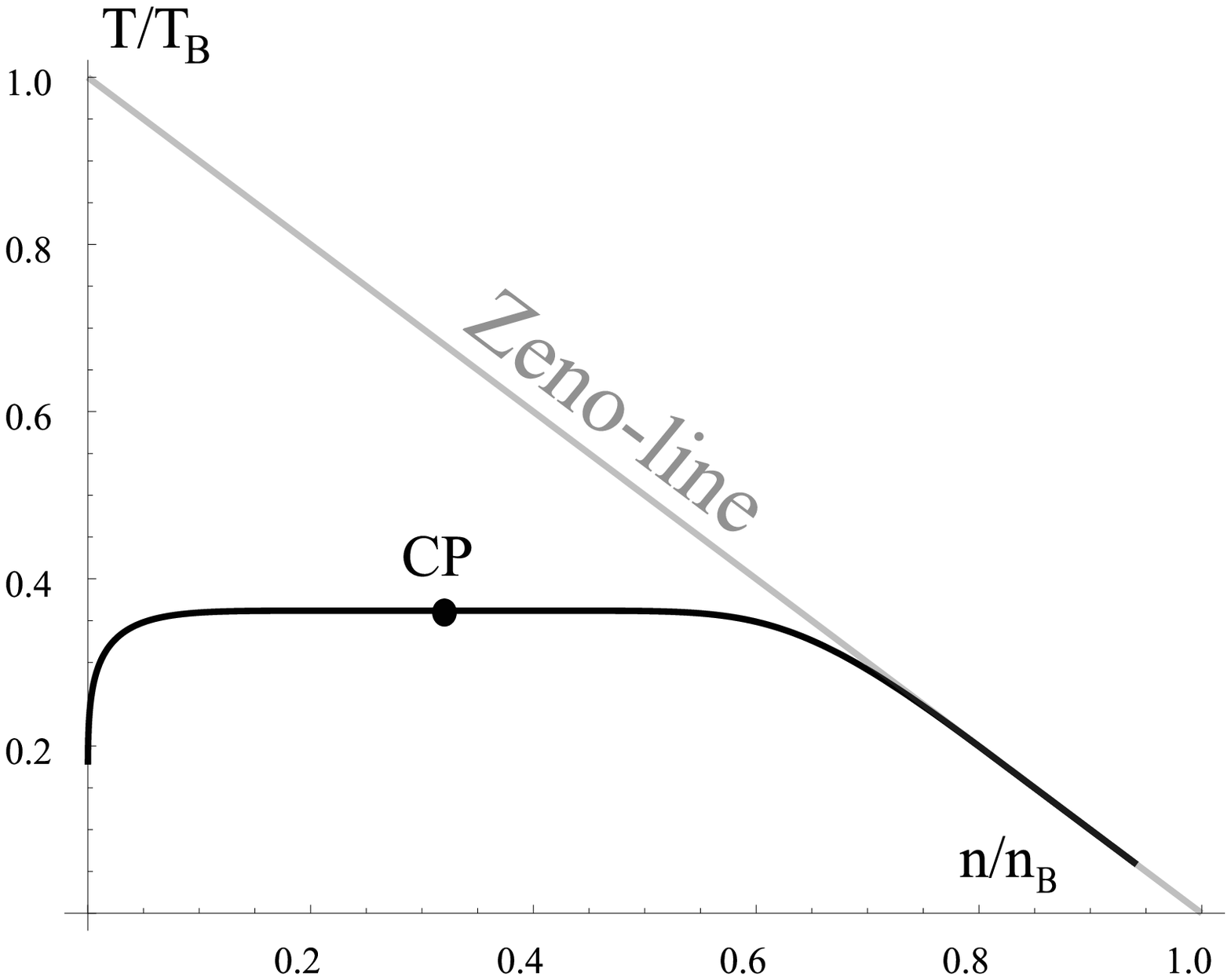}
  \caption{Binodal for 2D Lenard-Jones fluid with $\Phi_{attr}(r) \sim r^{-6}$ calculated as the isomorphic image of the binodal for $2D$ Lattice Gas (Ising model) in accordance with \eqref{projtransfr_nx}.}\label{fig_ising2d}
\end{figure}


The existence of global
isomorphism also provides the explanation of the fact about the cubic form of the binodal for real molecular fluids.  The fact that the shape of the binodal is almost cubic and characterized by the effective exponent
$\beta_{eff} = \frac{\partial\, \ln{\varphi}}{\partial\, \ln \tau}\approx 1/3\,,\varphi = \f{n_{l} - n_{g}}{n_c}$ in a wide temperature interval for
the system with short range interactions  was discovered long ago by J. E. Verschaffelt
\cite{crit_cubiccoex_leiden1896,eos_pitzer_purapplchem1989,
crit_sengersexp_physa1976} (see also  \cite{pcs_guggenheim_jcp1945}) and since then has been confirmed by the direct numerical simulations \cite{crit_simulreviewpanag_molphys1992,crit_ljfluid_pre1995}. In particular in \cite{crit_martynov_pre2009} the principle possibility to derive the nonclassical exponents for molecular liquids basing on the specific closure of the Ornstein-Zernike equation.

The proposed global isomorphism between LJ fluids and LG leads to the conclusion that the phase coexistence region of the latter is nothing but isomorphic image of the corresponding region of the phase diagram of the LG. For example applying the isomorphism transformation \eqref{projtransfr_nx} to the known result for 2D Ising model \cite{crit_2disingonsager_nuovochim1949}:
\begin{equation}\label{isingbinodal2}
  x = 1/2\pm f(t)^{1/8}\,,\quad f(t) = 1-\f{1}{\sinh^4\left(\,2J/t \,\right)}
\end{equation}
the binodal for 2D LJ fluid can be easily obtained (see Fig.~\ref{fig_ising2d}) and compared with the results of the computer simulations \cite{crit_lj2dim_jcp1991}. Also the inference that the cubic character of the binodal for real systems is the consequence of the cubic character of the binodal for the Ising model. So one can expect that the coexistence curve of the 3D Ising model is described in the ``zeroth approximation`` by the algebraic cubic curve in analogy with \eqref{isingbinodal2}:
\begin{equation}\label{isingbinodal3}
x = 1/2 \pm f(t)^{1/3}\Rightarrow t(x) = f^{-1}\left(\,\left|\,2x-1 \,\right|^3\,\right)\,,\quad f(1)=0
\end{equation}
where $f(t)$ is analytic function of the temperature.

In particular the isomorphism transformation \eqref{projtransfr_nx} applied to \eqref{isingbinodal3} allows to connect the critical amplitude $B_0$ for the order parameter of the LJ liquid:
\begin{equation}\label{binodal_lj}\overline{}
\f{n-n_c}{n_c} = \pm\,B^{\text{(LJ)}}_0\,|\tau|^{\beta}+\ldots\,,
\end{equation}
with that of the Ising model:
\begin{equation}\label{binodal_is}
2x-1 = \pm\,B^{\text{(Is)}}_0\,|\,\tau\,|^{\beta}+\ldots\,.
\end{equation}

Taking into account the relation \eqref{projtransfr_nx} between temperature variables $T$ and $t$ via comparison \eqref{binodal_lj} and \eqref{binodal_is} we obtain the following relation between amplitudes \[B^{(LJ)}_{0} \approx B^{(Is)}_0\,\left(\,1+z\,\right)^{\beta}\,.\]
As the amplitude $B^{(Is)}_0$ is known form exact result at $D=2$ or from computer simulation for $D=3$ (see e.g. \cite{crit_3disingmc_jmathphys1996}) one get the critical amplitude for LJ fluid. For the real $3D$ fluids this gives $B_0^{(LJ)}\approx 1.69$ for $2D$ case and $B_0^{(LJ)}\approx 1.85$ for $3D$ case.
Then the proximity of the numerical values of the critical exponents for the systems of the Ising model universality class to the rational numbers is quite natural. They do not necessary close to the mean-field values of the Landau theory and for the systems with the short-range interactions the mean-field behavior is not observed far away from the CP \cite{crit_2disingfss_pre1993}. This gives the grounds for the universal global cubic law for the binodal proposed in \cite{eos_zenoapfelbaum_jchempb2008}. In essential, the same global cubic law is widely used in computer simulations of the system with short ranged interactions \cite{crit_3disingmc_jmathphys1996,crit_simulmixture_prl2006}.
The fluctuations renormalize these rational exponents near the CP through small exponents $\eta$ and $\alpha$, which
determine the behavior of the correlation functions of the
density and the entropy in the immediate vicinity of the CP \cite{book_patpokr}. The fluctuations are responsible for the deviation from the linear diameter law \eqref{densdiam}. If we assume that the transformations between thermodynamic averages like $x$ and $n$ are generated by the transformations between the corresponding microscopic fields then such deviation is interpreted as the consequence of the nonlinearity of the transformation for the field variables. Indeed for the nonlinear
functional $F$ of the thermodynamic field $n(\m{r})$ in the presence of strong fluctuations $\left\langle\, F[n(\m{r})] \,\right\rangle \ne
F(\left\langle\, n(\m{r}) \,\right\rangle)$, where $
\left\langle\, \ldots \,\right\rangle
$ denotes the statistical average. From this point of view the transformation \eqref{projtransfr_nx} can be considered as the ``mean-field`` analog of commonly used linear mixing of the fields in the vicinity of the CP \cite{crit_rehrmermin_pra1973,crit_scalfieldsbrucewild_prl1992}.

\section{Conclusions}
In concluding Section we summarize the main results and discuss possible restrictions of the proposed approach.

The whole set of phenomenological facts about the interrelations between Zeno line, the rectilinear diameter, the binodal and the locus of the critical point revealed in \cite{eos_zenoapfelbaum_jcp2004,
eos_zenoapfelbaum_jchemp2006,zenoline_potentials_jcp2009,
eos_zenoapfelbaum1_jpcb2009} needs in unifying view from the microscopic approach of Statistical Mechanics. The proposed isomorphism allows to search such view in terms of the correspondence between the Hamiltonians of the LG and the fluids with short-ranged pair potentials. In particular it seems natural to connect the thermodynamic potentials for these systems. This allows to determine the parameters of mixing for the fields of the order parameter and the temperature \cite{crit_scalfieldsbrucewild_prl1992}. The results obtained for the locus of the critical point of LJ fluid and critical amplitudes of the equation of state show good correspondence with the results of computer simulations. Though the attractive part of the potential was taken into account in proposed version of the scale invariant mean-field approach the relations Eq.~\eqref{cp_simple},\eqref{tcphi} can be generalized to include the repulsive part of the potential. E.g. instead of \eqref{tcphi} the following relation arise:
\[
  \f{d\,\ln \left(\,T_c/T_Z \,\right) }{d\,\ln n_c/n_b} =   \f{d\,\ln \Phi_{\text{attr}}(n^{-1/d}_c) }{d\,\ln n_c/n_b}
\]
This relation along with the first equation in \eqref{cp_fluid} would give the closed system of equations to determine both $n_c$ and $z$ consistently. For the considered case of the LJ potential the repulsive part gives negligible correction so we neglect it here. But in general case the influence of the repulsive part on the locus of the critical point should be taken into account.

Note that the isomorphism relies on simple geometrical facts among which the tangency of the Zeno-line to the extrapolated liquid-vapor binodal into hypothetic region $T\to 0$ is the most crucial \cite{eos_zenoapfelbaum_jchempb2008}. For real systems the liquid branch is restricted by the triple point and this causes the problem of the correct determination of the parameter $n_b$. Obviously the slope of the Zeno line at $n_b$ influence essentially the position of the opposite Zeno state with $n\to 0$. To overcome this difficulty the correct determination of the limiting Zeno state $(T_Z,0)$ was proposed along with the determination of $n_b$ form the constraint $T_Z = a/b$  where $a$ and $b$ are the corresponding van der Waals constants for the equation of states. In principle it is possible to determine the values of $n_b$ using the data for the locus of the triple point and the slope of the binodal using the isomorphism transformation which allows to relate this data with the ones for the LG binodal. This will be the subject for further study.

It is interesting to note that in the limit $\varepsilon \to -d$, i.e. when the range of the interaction becomes infinite, $T_c \to T_B$. Thus the coexistence curve degenerates into triangle. This situation can be compared with the phase diagram for the infinite diluted polymer with the Flory theta-point \cite{flory_principles} (see also in  \cite{crit_coulombfisher_jsp1994}).

\end{document}